\documentstyle[11pt,paspconf,epsf]{article}
 
\begin{document}

\setlength{\unitlength}{0.75mm} 
 
\title{Determination of broadening functions using the Singular Value
Decomposition (SVD) technique}

\author{Slavek Rucinski}
\affil{Canada-France-Hawaii Telescope Co. Kamuela, HI 96743, USA}
 
\begin{abstract}
Cross-correlation function (CCF) has become the standard tool for  
extraction of radial-velocity and broadening information 
from high resolution spectra. It permits 
integration of information which is common to many spectral lines into one 
function which is easy to calculate, visualize and interpret. However,  
CCF is not the best tool for many applications where it should be
replaced by the proper broadening function (BF). Typical applications 
requiring use of the BF's rather than CCF's involve 
finding locations of star spots, studies of projected shapes of highly
distorted stars such as contact binaries (as no assumptions can be made about 
BF symmetry or even continuity) and [Fe/H] metallicity determinations 
(good baselines and avoidance of negative lobes are essential). 
It is stressed that the CCF's are not broadening functions.
The note concentrates on the advantages of determining the BF's through the 
process of linear inversion, preferably accomplished using the Singular Value 
Decomposition (SVD). Some basic examples of numerical operations are given in 
the IDL programming language.
\end{abstract}


\section{Convolution and cross-correlation}

Convolution is an operation that the nature does for us. 
We seldom see ``naked'' functions. These could be a convolution 
of a spectrum with the spectrograph's instrumental profile or with the radial 
component of the micro-turbulence velocity field in the 
stellar atmosphere or with a broadening function due to rapid
rotation of a star. Thus, instead of a function $f(u)$, we observe 
a function $h(x)$ which is a convolution with some other broadening 
function (BF), $g(x)$: 
\[ h(x) = \int_{-\infty}^{+\infty} f(u) \; g(x-u)\; du = f(x) \ast g(x) \]
This natural process can be easily simulated in numerical packages
(examples in the IDL programming language are marked by a
command-line prompt {\tt IDL>}) either by a special operator: \\
{\tt IDL> h = convol(f,\,h) } \\
or through the Fourier-transform multiplication and its inverse: \\
{\tt IDL> h = float(fft(fft(f,-1)*fft(g,-1),+1)) }

Cross-correlation is an operation which for real functions differs from the 
convolution really only in the symmetry of the arguments. 
For complex functions things are slightly different 
(real and imaginary parts have different symmetries), but astronomers 
observe real spectra so we do not have to worry about the 
mathematical nuances. 
\[ c(x) = \int_{-\infty}^{+\infty} f(u) \; g(u+x)\; du = f(x) \star g(x) \]
The cross-correlation (note the different asterisk above)
function can be computed numerically using: \\
{\tt IDL> lag = findgen(201) - 100 } \\
{\tt IDL> c = c\_correlate(f,g,lag) }

\section{Broadening functions}

Suppose we observe a sharp-line ($S(\lambda)$) and a broad-line ($P(\lambda)$) 
spectra and we want to determine the broadening and any other 
differences which make the latter spectrum more interesting 
than the former. The function $B(\lambda)$ can be a rotational 
broadening function for a single, rapidly-rotating star, or a more 
complex profile for two components of a binary, or for a star with
spots (where they would show as indentations in the function). 
The sharp-line spectrum $S(\lambda)$ is not 
free of some broadening. This can be the thermal broadening 
of lines or micro-turbulence or some other mechanisms; 
we call them jointly $T(\lambda)$. 
Thus, schematically, the sharp-line spectrum can be written as:
\[ S(\lambda) = (\sum_i a_i \delta(\lambda_i)) \ast T(\lambda) \]
while the broad-line spectrum, broadened additionally by $B(\lambda)$, can be 
written as:
\[ P(\lambda) = S(\lambda) \ast B(\lambda) =  (\sum_i a_i \delta(\lambda_i))
\ast T(\lambda) \ast B(\lambda) \]

The cross-correlation (CCF) with the sharp-line spectrum is frequently
taken as an estimate of $B(\lambda)$:
\[C(\lambda) = S(\lambda) \star P(\lambda) = 
S(\lambda) \star (S(\lambda) \ast B(\lambda)) = 
T(\lambda) \ast T(\lambda) \ast B(\lambda) \]
The new function ${\cal B}(\lambda)$, is {\it not 
identical\/} to $B(\lambda)$, because it inherits the common broadening
components (such as thermal, micro-turbulence, instrumental) from both 
spectra. Tonry  \& Davis (1979) showed that if those 
common components are represented 
by Gaussians, the addition is quadratical, which for these functions 
means repeated convolutions. 
Thus, CCF cannot be really used to replace the broadening function. 
But it can 
give us some approximation of it and will remain a useful tool to have some 
preliminary estimate on the degree of the line broadening. For symmetrical 
broadening functions, it will remain the simplest tool to determine the radial 
velocities simultaneously from many spectral lines.

\begin{figure}
\plottwo{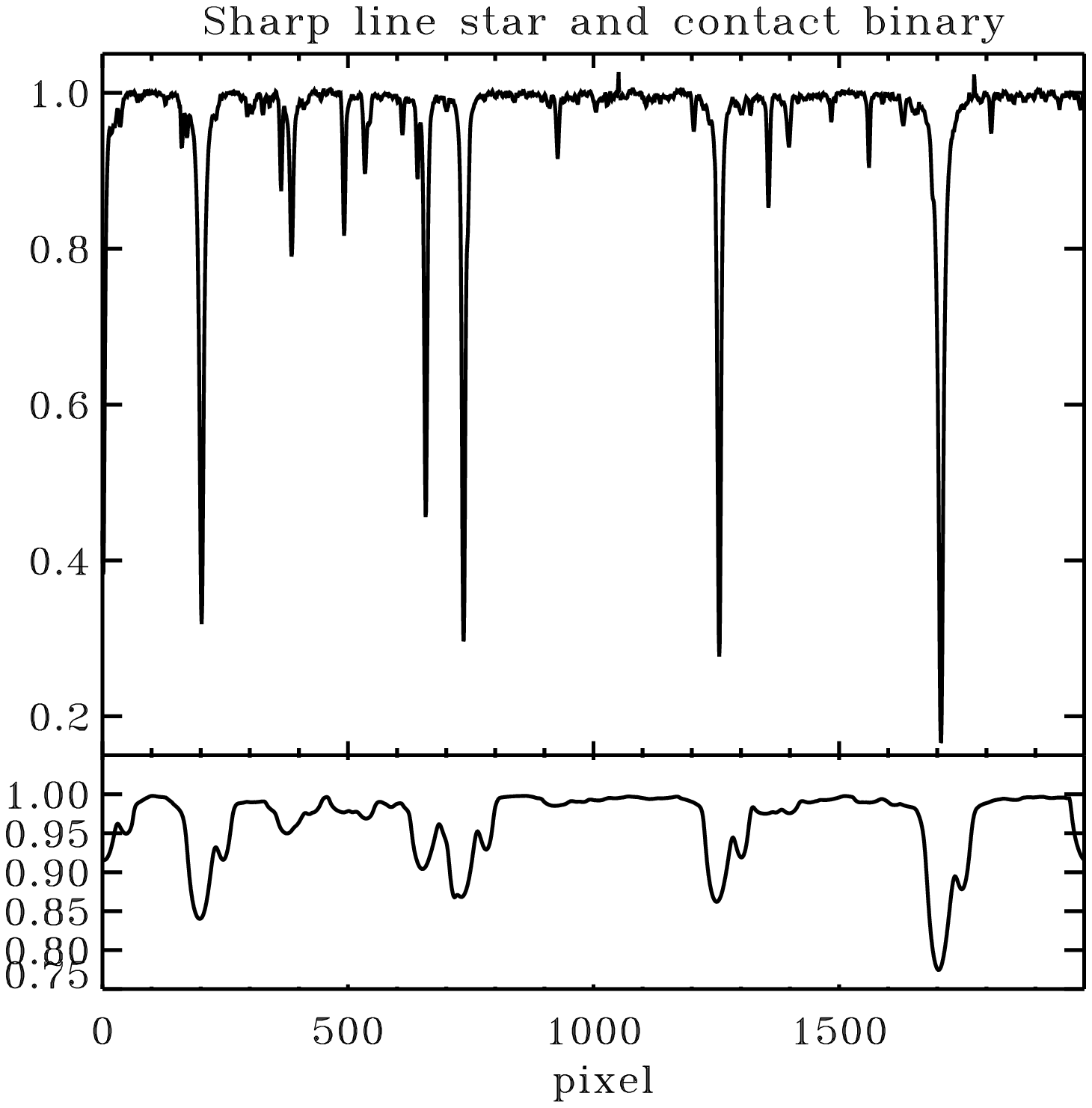}{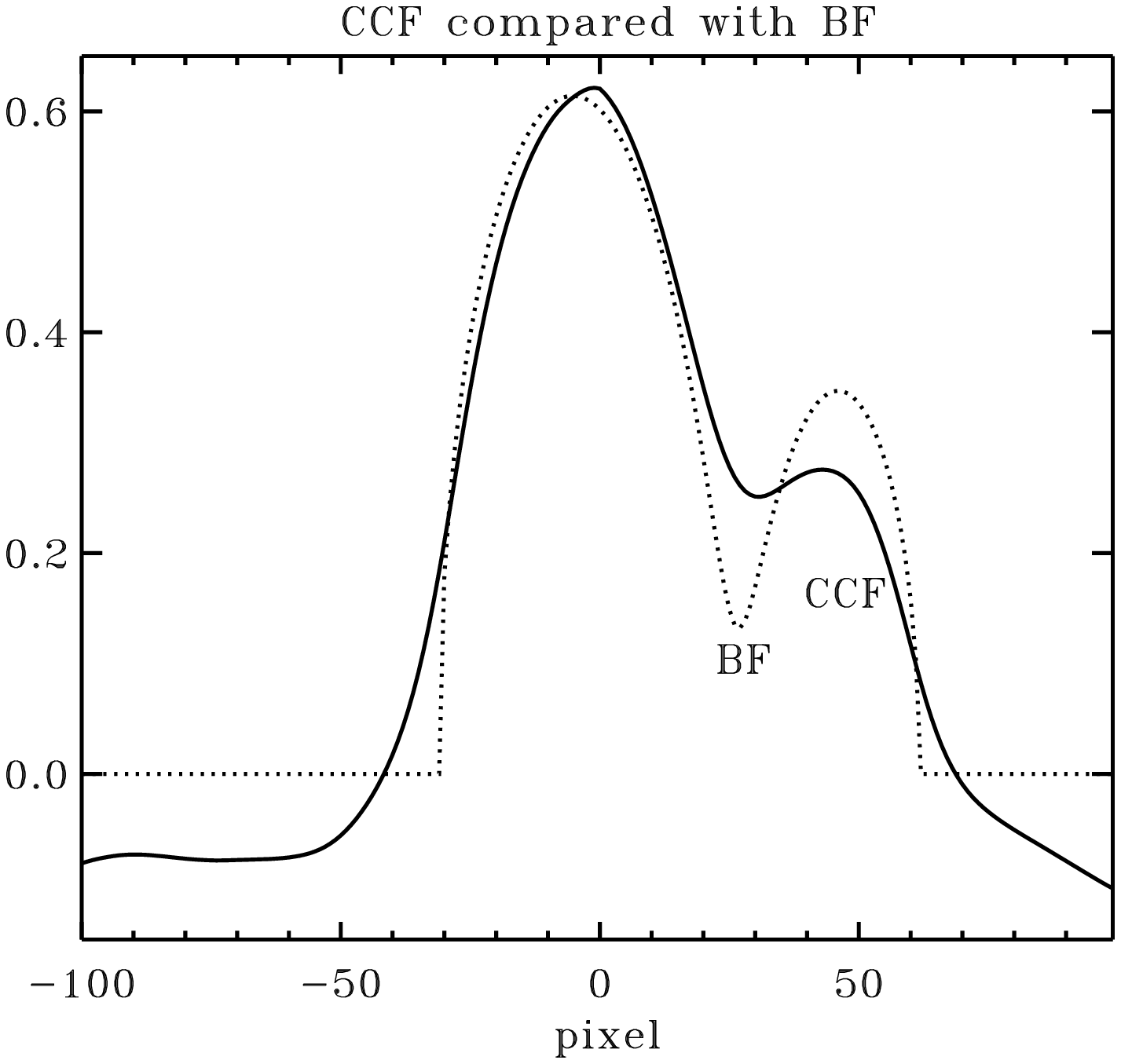}
\caption{The left panel shows a sharp-line spectrum (upper part) 
and the result of convolving it with the broadening function for 
a contact binary (lower part); this function is shown in the right 
panel by the dotted line (BF). The cross correlation of the broadened 
spectrum (with added noise to have $S/N = 100$) with the sharp 
spectrum is shown in the right panel by the continuous line. 
Notice the negative baseline and the peak-pulling in the CCF. 
} 
\end{figure}

The differences between the BF and the CCF can be seen when an artificial
broadened spectrum is created by a convolution and then the resulting 
spectrum is subject to the CCF operation. The result (Figure~1) is 
obviously different from the BF: 
The CCF shows negative baseline excursions and, most 
worryingly, it shows the ``peak-pulling'' effect which would lead to an 
under-estimate of the individual component velocities. While this last 
problem can be overcome by applying the TODCOR technique (Zucker \& Mazeh
1994), we clearly see that the CCF is not the BF.

\section{The Fourier transform de-convolution}

Some attempts to determine the broadening functions (Anderson et al.\ 1983) 
utilized the well known property of the Fourier transforms of a correspondence 
between convolutions and multiplications in the two relevant domains. Thus, a 
convolution: 
\[ P(\lambda)  = S(\lambda) \ast B(\lambda) \]
transformed with the Fourier transform $\cal F$ changes into a product of the 
transforms:
\[ {\cal F} \left\{ P(\lambda) \right\} = 
   {\cal F} \left\{ S(\lambda) \right\} \cdot 
   {\cal F} \left\{ B(\lambda) \right\} \]

\noindent
Therefore, the broadening function can be restored with:
\[ B(\lambda) \simeq {\cal F}^{-1} \left\{ {\cal F} 
 \left\{ P(\lambda) \right\} /
 {\cal F} \left\{ S(\lambda) \right\} \right\} \]

\noindent
This can be done in a compact way with: \\
{\tt IDL> b = float(fft(fft(p,-1)/fft(s,-1),+1))  }

While the mathematical background is simple and easy, the practice is just the 
opposite. First, the resulting $B(\lambda)$ spans the whole spectral window, 
so that one determines a lot of zeroes; there is no ``compression'' 
information whatsoever. But, more importantly, the division operation 
usually ... does not 
work: the high frequency noise becomes amplified and some sort of frequency 
filtering is needed. The result may then actually depend on the 
applied filter. 

Some authors (see for example the large collection of works of D.\ F.\ Grey) 
use spectra transformed into the frequency domain without the division 
step. This can be done for broadening mechanisms describable 
by simple functions or obeying some symmetries, but fails 
in cases of spots or of BF's of close binary systems.

\section{Convolution in the formalism of linear equations}

There are two main issues that re-casting convolution into a set of linear 
equations can resolve. These are: (1)~How to channel information 
over the whole spectrum (say 2000 pixel long) into the BF window 
(say 200 pixels long)?
(2)~How to utilize all information contained in sharp-line spectra and 
remove the influence of the noise in the continuum (which carries 
no information)? 

The convolution can be written as an over-determined system of 
linear equations 
which link a sharp-line spectrum $\vec{S}(n)$, via the broadening function 
$\vec{B}(m)$, with the broadened spectrum $\vec{P}(m)$. The mapping is through 
the ``design matrix'' $\widehat{Des}(m,n)$ which is formed from the 
sharp line spectrum $\vec{S}(n)$ by consecutive vertical shifts by one element.
In IDL, this can be done with a simple routine:

\noindent
{\tt function map4,s,m \\
; m - must be odd, n must be even \\
n = n\_elements(s) \& t = fltarr(m) \# fltarr(n-m+1) \\
; t(m,n-m) = t(small,large-small) dimensions \\
for j = 0,m-1 do for i = m/2,n-m/2-1 do t(j,i-m/2)=s(i-j+m/2) \\
return,t \\
end }

An example of using this routine to create a design matrix 
for a 201-pixel long window would be: {\tt IDL> des = map4(s,201)}. 
The program spectra must accordingly be trimmed to {\tt n-m+1} 
with: {\tt IDL> p = p(m/2:n-m/2-1)}.
The system of equations has a familiar form of the over-determined 
linear set:

\begin{center}
\begin{picture}(50,60)
\put(0,0){\framebox(15,60)[c]{$\widehat{Des}$}}
\put(20,22){\framebox(5,15)[c]{$\vec{B}$}}
\put(30,28){=}
\put(40,0){\framebox(5,60)[c]{$\vec{P}$}}
\end{picture}
\end{center}

\section{Singular value decomposition (SVD)}

One of the traditional ways of solving the system of equations above 
would be to transform it into a system of the ``normal'' equations of 
the size reduced from $m \times (n-m)$ to $m \times m$ by
multiplication of both sides by the transpose of the design matrix.
The result would be the BF defined in the least-squares sense, and
it is possible to stop at this point. However, the Singular Value 
Decomposition technique also gives us such an answer, but -- in addition --
makes it possible to remove the influence of the continuum and its noise.   

The SVD technique is beautifully described in the ``numerical techniques 
Bible'' of Press et al.\ (1986). They present it as a somewhat magic 
black box and for most users it is just fine. If you want to 
learn how the technique really works, then the books of Golub \& Van Loan
(1989) or Craig \& Brown (1986) are probably the best references. 

The essence of the SVD is the property that one can represent any matrix by a 
product of 3 matrices; in our case: $\widehat{Des} = \widehat{U} \,\widehat{W} 
\, \widehat{V}^T$. These matrices are: the column ortho-normal 
$\widehat{U}$ and 
$\widehat{V}$ and the diagonal matrix $\widehat{W}$ (this is 
really a vector containing the diagonal elements). The property of 
the columns in $\widehat{U}$ and $\widehat{V}^T$ is that the 
following products, $\widehat{U}^T \widehat{U} = 
\widehat{I}$ and $\widehat{V}^T \widehat{V} = \widehat{I}$, 
give the unity array $\widehat{I}$ (1 on the diagonal).

\begin{center}
\begin{picture}(80,60)
\put(0,0){\framebox(15,60)[c]{$\widehat{Des}$}}
\put(18,28){=}
\put(25,0){\framebox(15,60)[c]{$\widehat{U}$}}
\put(45,22){\framebox(15,15)[c]{$\widehat{W}$}}
\put(65,22){\framebox(15,15)[c]{$\widehat{V}^T$}}
\end{picture}
\end{center}

\noindent
In IDL, the operation is represented by: {\tt IDL> svdc,des,w,u,v,/double}.
Here, {\tt des} is the only input quantity and the remaining
parameters are what the routine produces as 
output. The keyword {\tt /double} is for higher precision and is optional.

One can check the correctness of the operations  by the following 
commands: \\
{\tt IDL> wf = fltarr(m,m) }\\
{\tt IDL> for i = 0,m-1 do wf(i,i) = w(i) }\\
{\tt IDL> des\_check = u \#\# wf \#\# transpose(v) }

The three new matrices are all invertible: 
$\widehat{U}$ and $\widehat{V}$ are ortho-normal arrays, so that 
their inverses are just transposes, while the diagonal array 
$\widehat{W}$ is replaced by a diagonal array $\widehat{W1}$, 
with the diagonal elements containing the inverses, 
$w1_{i,i} = 1/w_i$: \\
{\tt IDL> w1 = fltarr(m,m) } \\
{\tt IDL> for i = 0,m-1 do w1(i,i) = 1./w(i) } \\

\noindent
The solution is given by: 
$ \vec{B} = \widehat{V} \widehat{W_1} (\widehat{U}^T \vec{P}) $ or
schematically:

\begin{center}
\begin{picture}(140,60)
\put(0,22){\framebox(5,15)[c]{$\vec{B}$}}
\put(8,28){=}
\put(16,22){\framebox(15,15)[c]{$\widehat{V}$}}
\put(34,22){\framebox(15,15)[c]{$\widehat{W1}$}}
\put(52,22){\framebox(60,15)[c]{$\widehat{U}^T$}}
\put(118,0){\framebox(5,60)[c]{$\vec{P}$}}
\end{picture}
\end{center}
 
Numerically: 
{\tt IDL> b = reform(v\#\#w1\#\#(transpose(u)\#\#p))}. A routine
makes this simpler: {\tt IDL> b = svsol(u,w,v,p,/double)}. 
The elements of $\vec{B}$ are all independent, so that any -- even strange or
discontinuous -- broadening functions can be restored as no condition imposed
on the smoothness or symmetry of the result. Note that, if only
one sharp-line template is used, the decomposition operation {\tt svdc}
is done only once, for possibly many broad-line spectra {\tt p},
each giving a separate solution {\tt b}.

\section{Advantages and disadvantages of the SVD approach}

On the positive side: (1)~The problem can be treated as a 
linear-equations. (2)~An ``inverse'' of the {\it rectangular\/} 
array $\widehat{Des}$ is possible. (3)~The solution of $\vec{B}$ 
is defined in the least-squares sense 
(shortest modulus). (4)~The result is the real broadening function.
But there are also minuses: (5)~One must solve a large system 
of, say, 2000 equations for 200 unknowns. (6)~One must know a priori 
how many unknowns. (7)~Initially, the results may turn out quite 
poor, because of the presence of 
plenty of linearly-dependent equations in the system (parts of spectra 
where the featureless continuum provides no broadening information).

The SVD approach offers a simple resolution of (7)
as {\it it permits removal of the effects of the continuum 
in an objective way}. The key element here are the singular values 
contained in the diagonal of $\widehat{W}$. Since the solution 
involves $1/w_i$, small values in $w_i$ spoil 
the solution. These are exactly those problematic values that one wants to 
avoid. Thus, by rejecting of small values of $w_i$, one can (i) remove the 
linearly dependent equations, 
(ii) diminish the influence of the noise from the 
continuum, (iii) reduce the influence of the computer round-off errors (which 
enter multiplied by the order of the problem) and 
(iv) reduce the number of the 
unknowns (because the system is usually not over-determined at all). All these 
properties are related to the ``conditioning'' of the array $\widehat{Des}$. 
The reader is directed to the source texts on this subject for 
further reading. 

\begin{figure}
\plottwo{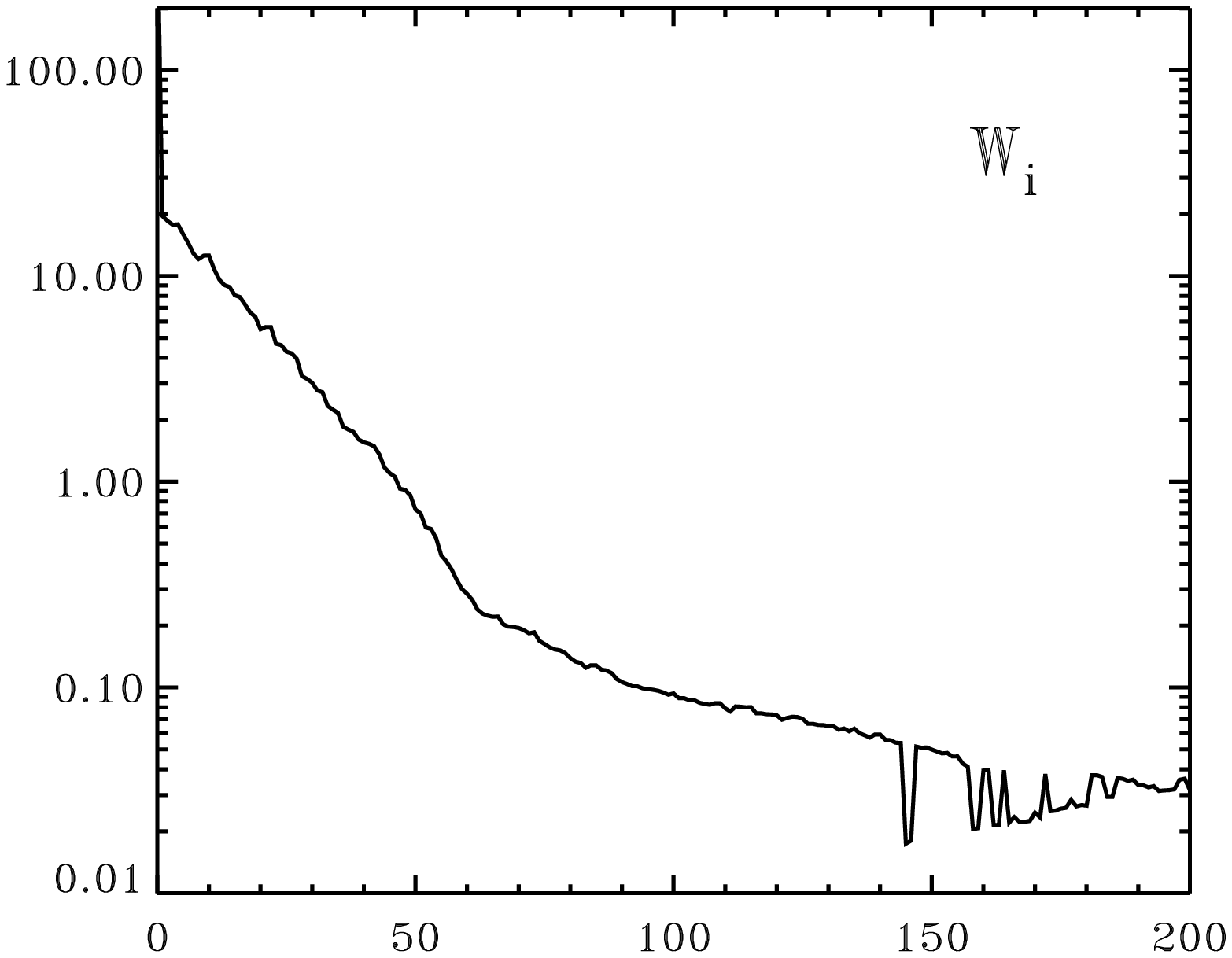}{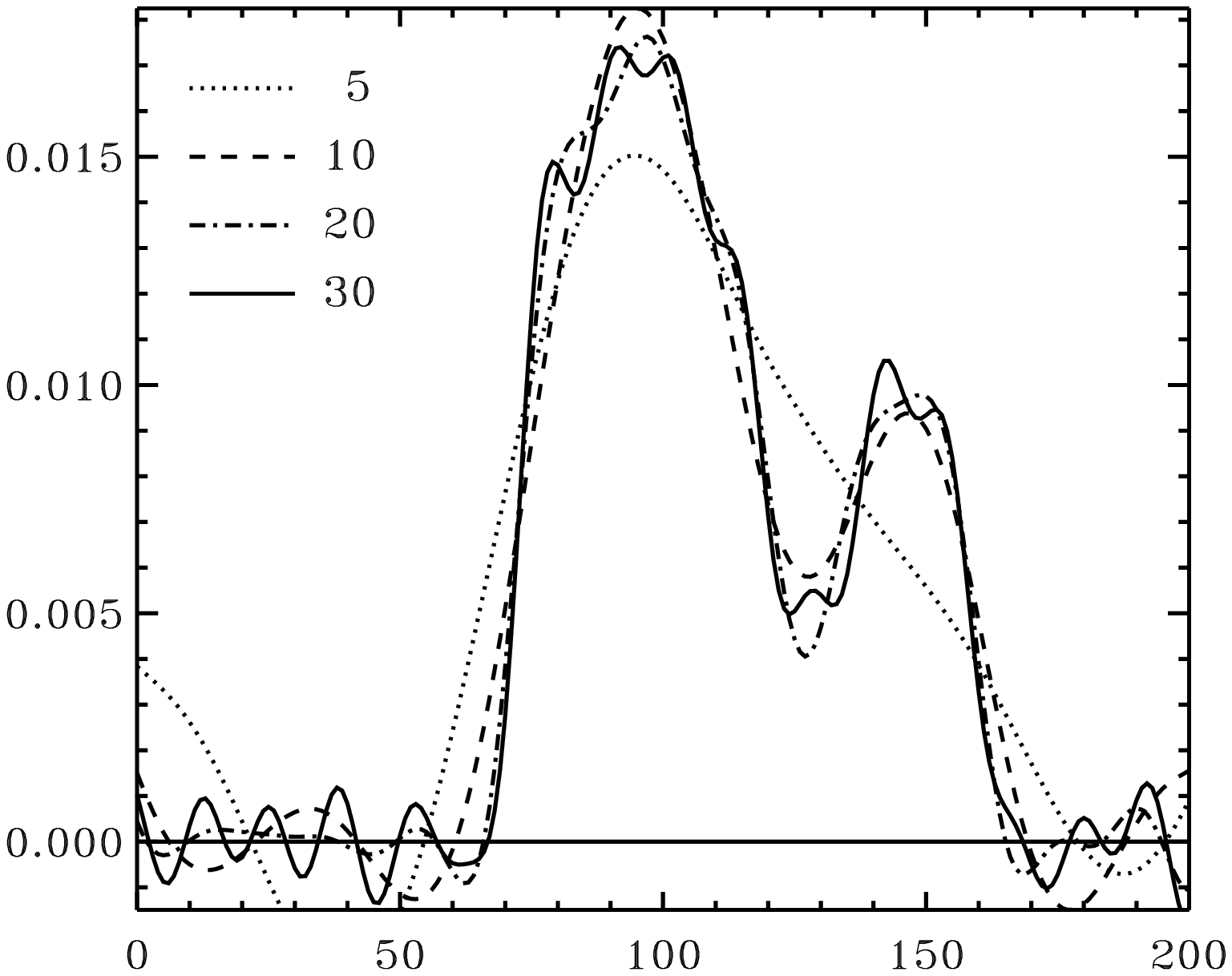}
\caption{The left panel shows the run of the singular values for 
the sharp-line spectrum in Figure~1. The right panel gives the 
solutions for the broad-line spectrum in the same figure (with 
added noise at $S/N = 100$) utilizing the first 5, 10, 20 and 30 
singular values.} 
\end{figure}

The important factor is max($w_i$)/min($w_i$) which provides an estimate 
on how many of the singular should be used. 
In practice, one can keep on adding more $w_i$ and 
see the successively better solutions. The diagonal arrays 
$\widehat{W}$ and  $\widehat{W1}$ will have then elements: 
$w_i = w_0, w_1, w_2, w_k, ... , w_{m-1}$ and 
$w1_i = 1/w_0, 1/w_1, 1/w_{k-1}, 0, ... , 0$ with $k$ 
(we call it the order of solution) spanning the whole range 0 to $m-1$.  
In IDL, this can be done by forming a square matrix of solutions:

\noindent
{\tt b = fltarr(m,m) \\
for i = 0,m-1 do begin \\
wb = fltarr(m) \\
wb(0:i) = w(0:i) ; first i+1 singular values used, rest zero \\
b(*,i) = svsol(u,wb,v,p,/double) \\ 
end for}

We note in passing, that the arrays $\widehat{U}$ and $\widehat{V}$ 
have very special properties as they contain the basis vectors in 
the spaces of the spectra and broadening functions, respectively. 
One can see this by analyzing a diagonalized system: 
$\widehat{W}\, \vec{Z} = \vec{D}$, obtained by keeping the same 
$\widehat{W}$, with $\vec{Z} = \widehat{V}^T \vec{B}$ and $\vec{D} 
= \widehat{U}^T \vec{P}$. The solution of the diagonalized 
system would be then: 
$\vec{Z} = \vec{D}/\vec{W}$, but, in practice, the diagonal 
of $\widehat{W}$ is a vector, so that $z_i = d_i/w_i$. 
Plotting the columns of $\widehat{U}$ and $\widehat{V}$ 
can tell one a lot about the conditioning of the solution.

\section{A few notes on the SVD solutions}

The first question is: How far in $k$ should one go and where to stop?
The essential operation is to plot (usually in log units) the 
vector $\vec{W}$ (Figure 2). There are 3 parts of it: (1)~the good, 
large singular values, (2)~the small values usually representing 
the noise in the spectrum $S(\lambda)$ 
and (3)~the numerical errors. You may want to stop 
no later than at the kink below the good part. 
But the real ``quality control'' is the fit to $\vec{P}$. 
If the error of the fit stops 
decreasing, you have found the right point (Figure 3). 
Beyond that point, you will start 
fitting the noise! However, it is useful to analyze the solutions 
for different orders $k$ and see how they first improve and then get worse. 
Sometimes the fit will remain poor, in spite of the leveling of the error 
curve; this usually means a wrong choice of the sharp-line spectrum.
The standard error of the fit can be calculated from: \\
{\tt sig = fltarr(m) ; error    \\
  pred = des \#\# transpose(b) ; predicted fits \\
  for i=0,m-1 do sig(i) = sqrt(total((pred(i,*)-p)\symbol{'136}2)/m)}

Usually, even very low order solutions are well defined (Figure 2), but 
stopping early is not always advisable because 
this leads to a loss of resolution, as then not all basis vectors contribute. 
The solution which -- in principle -- has all elements in $\vec{B}$
independent suddenly acquires inter-element correlations.
Thus, it may be advantageous to go to a highest possible order ($k=m$) and  
thus insure that elements of $\vec{B}$ are uncorrelated, and then 
decrease the noise by smoothing.

\begin{figure}
\plottwo{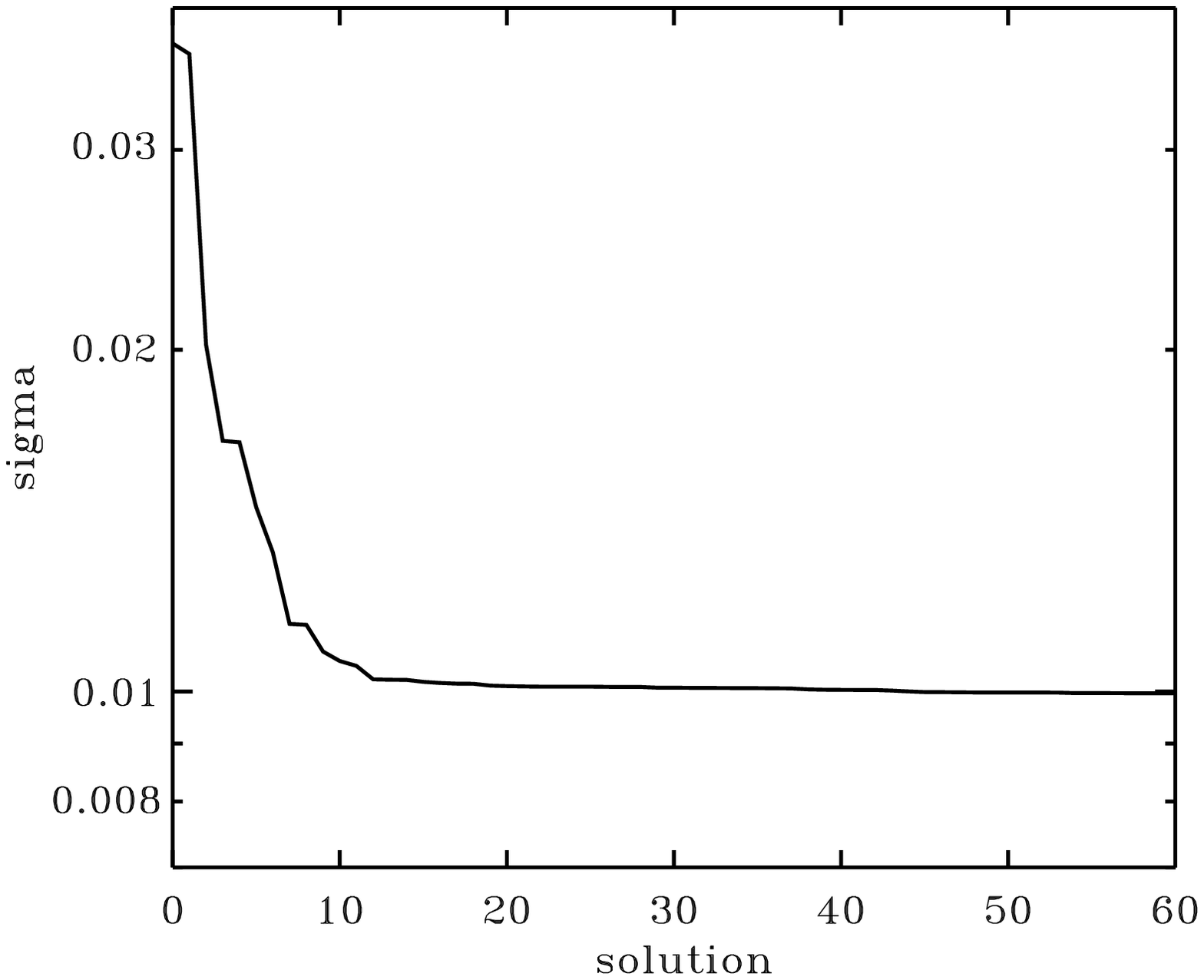}{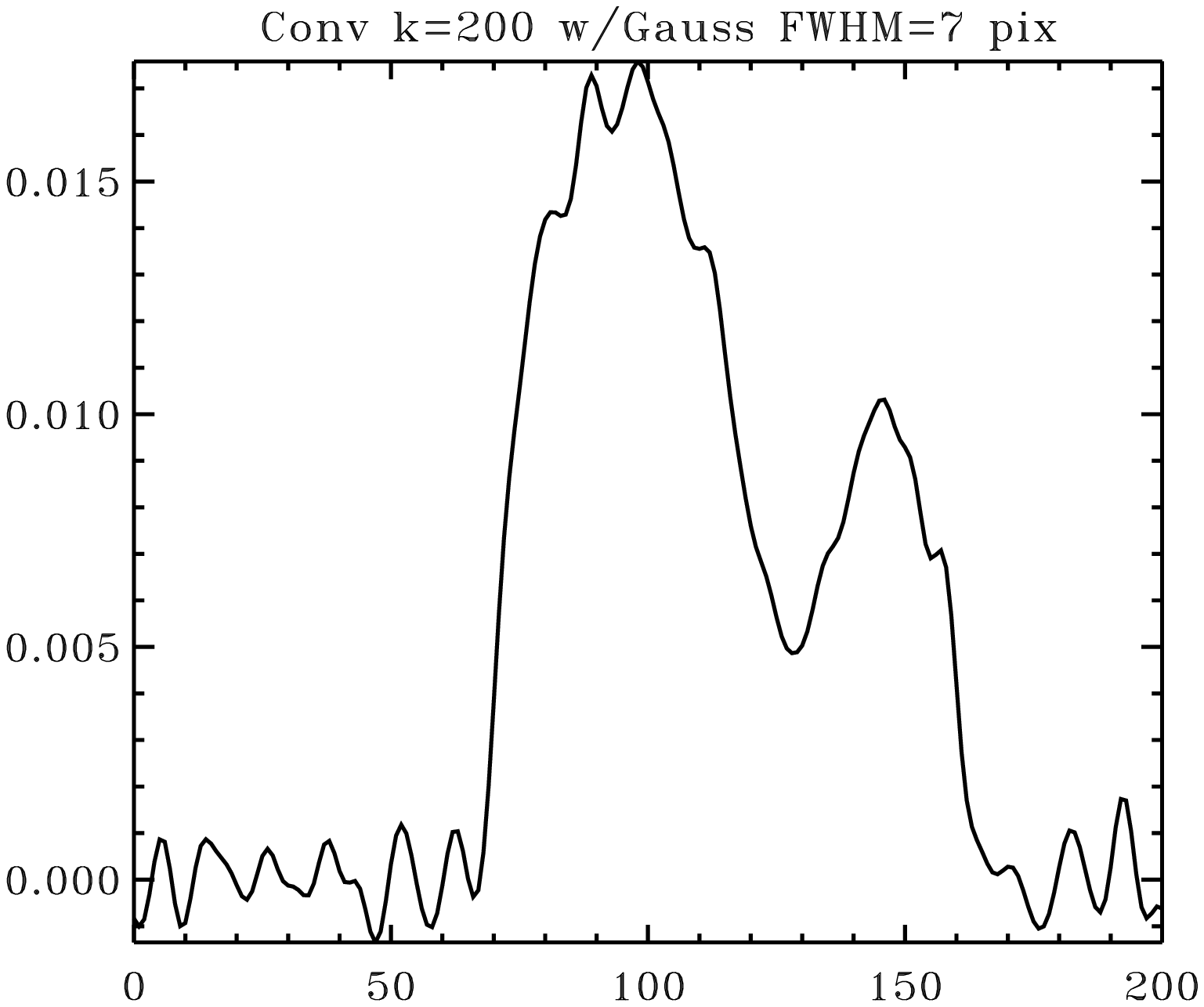}
\caption{The left panel shows the mean standard error of the fit 
for our example solutions utilizing progressively more singular 
values. It is obvious that the first 10 -- 15 singular values give 
an adequate fit (compare with Figure 2). 
The right panel shows the solution utilizing all 
200 singular values, convolved with a Gaussian. This approach may 
offer a better control over the final resolution of the BF.}
\end{figure}

One should be aware that the errors may be under-estimated 
for the case of truncated ($k < m$) solutions. While the prescriptions of 
Rix \& White (1992) and Rucinski, Lu \& Shi (1993) are based on the 
theory of the full SVD, the error analysis for the truncated case 
has not yet been done. In this situation, it may be advantageous to 
utilize techniques of the external estimates, such as the bootstrap or Monte 
Carlo. This subject certainly requires more work ...

\section{Conclusions}

One can position the SVD technique of linear broadening 
function restoration as 
located between the cross-correlation and the direct modeling of the spectra. 
The BF's determined with it are much better defined than the 
CCF's, and are true 
broadening functions, not their proxies. They also integrate the geometrical 
information from a spectrum, but give well defined baselines without the CCF's 
negative fringes. They do require a bit more computer work, 
but several numerical packages can easily handle large linear 
systems of equations involved here. Obviously, they 
cannot replace the spectrum synthesis, 
if these are needed, but can be a useful tool in their preparation.

\end{document}